# NIR imaging and modeling of the core of M100

J.H. Knapen[1], R.F. Peletier[2,3], I. Shlosman[4], J.E. Beckman[3], C.H. Heller[5] & R.S. de Jong[6]

[1] Université de Montréal, Dép. de Physique, C.P. 6128, Succ. Centre Ville, Montréal, Québec, H3C 3J7 Canada; and Observatoire de Mont Mégantic
[2] Kapteyn Institute, Postbus 800, NL-9700 AV Groningen, the Netherlands
[3] Instituto de Astrofísica de Canarias, E-38200 La Laguna, Tenerife, Spain
[4] Dept. of Physics and Astronomy, Univ. of Kentucky, Lexington, KY 40506-0055
[5] Univ. Sternwarte, Geismarlandstraße 11, D-37083 Göttingen, Germany
[6] Univ. of Durham, Physics Dept., South Road, Durham DH1 3LE, UK

**Abstract.** High-resolution NIR and optical images are used to constrain a dynamical model of the circumnuclear star forming (SF) region in the barred galaxy M100 (=NGC 4321). Subarcsecond resolution allowed us to distinguish important morphological details which are easily misinterpreted when using images at lower resolution. Small leading arms observed in our $K$-band image of the nuclear region are reproduced in the gas flow in our model, and lead us to believe that part of the $K$ light comes from young stars, which trace the gas flow.

In the optical, the central region of M100 shows a set of tightly wound star-forming spiral armlets. The arms can be traced outward through the bar of the galaxy and connect to the main spiral arm pair in the disk. They are accompanied by well-defined dust lanes (Knapen et al. 1995a). In the NIR however the central region looks markedly different. In the $0''\!.8$ $2.2\mu$m $K$-image we obtained at UKIRT (Knapen et al. 1995a), one distinguishes, from inside out, a small bulge; an inner barlike region, with identical position angle and ellipticity to the large-scale bar; two leading armlets; two symmetric emission peaks; and an oval ring-like zone where the SF armlets are found in the optical. The $K$ contours in this zone are smooth and show hardly any indication of spiral arms or dust lanes. This change in morphology from optical to NIR is caused by different distributions of the stellar populations, along with the much reduced absorption by dust at $2.2\mu$. The two symmetrically placed peaks of $K$ emission are massive starburst regions of roughly the same age (Knapen et al. 1995a), where at least part of the $K$ emission is likely to come from young stars (e.g. O stars and K supergiants). The fact that the $K$ isophotes become progressively elongated and skewed towards the position angle of the bar both outside *and* inside the "ring" is a strong indicator in favor of a double inner Lindblad resonance in this galaxy.

The need for high ($< 1''$) resolution NIR imaging in this work becomes especially clear when comparing the results from our work with those from two recent papers, where the morphology of the inner $\sim 15''$ region is misinterpreted from $K$-band images with resolutions of $1''\!.9 \pm 0''\!.2$ (Shaw et al. 1995), and $2''\!.9$ (Sakamoto et al. 1995). Both authors claim the existence of a $\sim 10''$ radius nuclear bar (already reported by Pierce 1986), whereas we resolved this structure into an inner part of the bar of some $5''$, the leading arms, and two symmetric peaks of $K$ emission (Knapen et al. 1995a). As a consequence, only our high-resolution $K$ imaging shows that the inner and outer parts of the bar are aligned, and not at different position angles such as claimed by the other authors.

We have modeled the stellar and gas dynamical processes of the nuclear ring-like structure and associated features by means of 3D numerical simulations using a method described by Heller & Shlosman (1994). Non-linear orbit analysis was used to verify the positions of the ILRs by determining the spatial extent of the family of orbits oriented along the minor axis of the bar. Our modeling shows explicitly that the dominant morphology in the center of M100 can be explained by the gas response to the stellar bar potential (Knapen et al. 1995b). We find that a system of trailing and leading shocks in the gas in the vicinity of the ILRs shows a robust behavior. We have been able to identify and model the main regions of SF there, corresponding to four compression zones. Two zones of SF correspond to the so-called "twin peaks," and two additional ones are found where a pair of large-scale trailing shocks interacts with a pair of leading shocks. Young massive stars ($\sim 10^7$ yrs and less) are present in the resonance region in addition to the old population. Emission from young stars, which trace the gas flow, and from hot dust can explains why we see the leading arms in $K$ emission, whereas they are theoretically expected to occur in the gas.